\documentstyle[aps,prl,twocolumn,epsfig]{revtex}

\begin{document}

\twocolumn[\hsize\textwidth\columnwidth\hsize\csname
@twocolumnfalse\endcsname

\title{Epitaxial MgB$_2$ superconducting thin films with a transition
temperature of 39 Kelvin }
\author{W. N. Kang\cite{email}, Hyeong-Jin Kim, Eun-Mi Choi, C. U. Jung, and Sung-Ik Lee
}
\address{National Creative Research Initiative Center for Superconductivity\\
and Department of Physics, Pohang University of Science and Technology,\\
Pohang 790-784, Republic of Korea\\
\vskip 1pc
(Submitted March 2, 2001)}
\draft
\maketitle

\begin{abstract}

\end{abstract}

\vskip 0.5pc]

\bigskip

{\bf The recent discovery of the binary metallic MgB}$_{{\bf 2}}${\bf \
superconductor \cite{Nagamatsu} having a remarkably high transition
temperature (T}$_{{\bf c}}${\bf ) of 39 K has attracted great scientific
interest \cite{Kang1,Jung1,Takano,Finnemore,Kotegawa,Karapetrov,Bud'ko}.
With its metallic charge carrier density\cite{Kang1} and the strongly linked
nature of the inter-grains in a polycrystalline form \cite
{Labalestier,Canfield}, this material is expected to be a very promising
candidate for superconducting device \cite{Gupta} as well as large-scale
applications. However, the fabrication of thin films of this material has
not been reported yet. Here, we report the growth of high-quality epitaxial
MgB}$_{{\bf 2}}${\bf \ thin films on (1102) Al}$_{{\bf 2}}${\bf O3 and (001)
SrTiO}$_{{\bf 3}}${\bf \ substrates by using a pulsed laser deposition
technique. The thin films grown on Al}$_{{\bf 2}}${\bf O}$_{{\bf 3}}${\bf %
\ substrates show a T}$_{{\bf c}}${\bf \ of 39 K with a sharp
transition width of }$\sim ${\bf \ 0.7 K. The critical current
density in zero field is }${\bf \sim }$ {\bf 8 x 10}$^{{\bf
6}}${\bf \ A/cm}$^{{\bf 2}}${\bf \ at 5 K, suggesting that this
compound has great potential for industrial applications. Also,
Al}$_{2}${\bf O}$_{{\bf 3}}${\bf \ is a very promising substrate
in terms of superconducting device applications because of its
high thermal conductivity and small dielectric constant. For the
films deposited on Al}$_{{\bf 2}}${\bf O}$_{{\bf 3}}${\bf , X-ray
diffraction patterns indicate a c-axis-oriented crystal structure
perpendicular to the substrate surface whereas the films
deposited on SrTiO}$_{{\bf 3}}${\bf \ show \{101\} plane as the
preferred orientation.}

The growth technique used in this study was reported elsewhere.
MgB$_2$ thin films were deposited on Al$_{2}$O$_{3}$ (AO) and
SrTiO$_{3}$ (STO) substrates by using pulsed laser deposition.
The laser energy density was 20 - 30 J/cm$^{2}$ at a laser flux
of 600 mJ/pulse. The resistivity measurements were carried out
using the dc four-probe method. The dc magnetic properties were
measured with a Quantum Design MPMS superconducting quantum
interference device magnetometer. The structures were analyzed
using a x-ray diffractometer (XRD).

The typical temperature dependence of the resistivity of
MgB$_{2}$ grown on AO is shown in Fig. 1. The inset is a
magnified view near the T$_{c}$ region. The resistivity of the
thin film begins to enter the superconducting transition at 39 K
and goes to zero resistance at 37.6 K. A very sharp transition,
with a width of ${\bf \sim }$ 0.7 K from 90\% to 10\% of the
normal state resistivity, is evident in the inset. This is
comparable to most reported values for high-quality bulk samples
{\cite {Jung1,Takano,Finnemore,Kotegawa}}. The normal-state
resistivity at 290 K was ${\bf \sim }$ 4.7 $\mu $cm, indicating
an intermetallic nature with a relatively high charge carrier
density {\cite{Kang1}}. This resistivity is smaller than those
for polycrystalline MgB$_{2}$ wire {\cite{Canfield}} and for bulk
samples synthesized under high pressure {\cite {Jung1,Takano}}.
Most of our films fabricated under the same conditions showed a
similar superconducting transition around 39 K.

Fig. 2 shows the zero-field-cooled (ZFC) and the field-cooled (FC) dc
magnetization (M) curves of a MgB2 thin film in a 10 Oe field applied
parallel to the c-axis. The irreversibility temperature detected at 37.5 K
coincides with the zero-resistance temperature obtained from resistivity
measurement. The ZFC curve shows a rather broad diamagnetic transition
compared to the resistivity data. To estimate the critical current density (J%
$_{c}$), we measured the M-H loop for the same sample as a
function of temperature, as shown in the inset of Fig. 2. At 5 K,
the J$_{c}$ calculated using the Bean model was ${\bf \sim }$ 8 x
10$^{6}$ A/cm$^{2}$ at zero field and ${\bf \sim }$10$^{6}$
A/cm$^{2}$ at 1 T, which are 20 times higher than the values
obtained for a MgB$_{2}$ wire {\cite{Canfield}} by using
transport measurements and slightly smaller than the values for
Hg-based superconducting thin films {\cite{Krusin,Kang2}}. It
should be noted that we used the sample size rather than the
grain size in calculating J$_{c}$.

Structural analysis was carried out by XRD, as shown in Fig. 3. The a- and
the c-axis lattice constants determined from the (101) and the (00{\it l})
peaks are observed to be 0.310 and 0.352, respectively. Interestingly, the
XRD patterns indicate that the films deposited on (a) AO are epitaxially
aligned with the c axis while the films on (b) STO is well aligned with the
(101) direction normal to the substrate planes. These results suggest that
we may be able to control the orientation of MgB$_{2}$ thin film simply by
using different substrates if the optimum growth condition is explored.

We also find that the AO substrates are chemically very stable during the
heat treatment at high temperatures in Mg vapor. For the STO substrates,
however, the chemical reaction between the substrate and Mg was observed and
appeared as minor impurity phases, as shown in Fig. 3 (a). Considering that
STO has a cubic structure with a lattice constant of 0.389 nm and R-plane AO
has a$_{0} $ = 0.476 nm and b$_{0}^{\ast }$ = (c$_{0}^{2}$ + 3a$_{0}^{2}$)$%
^{1/2}$ = 1.538 nm, it is quite striking that MgB$_{2}$ thin films grow with
preferred orientations on AO and STO substrates even though the
lattice-matching relationship between the MgB$_{2}$ and the substrates is
not well satisfied. Further microscopic investigation of the MgB$_{2}$
crystallographic relations on AO and STO substrates is required.

\bigskip
\acknowledgments
This work is supported by the Ministry of Science and Technology of Korea
through the Creative Research Initiative Program.

\newpage
\begin{figure}[tbp]
\centering \epsfig{file=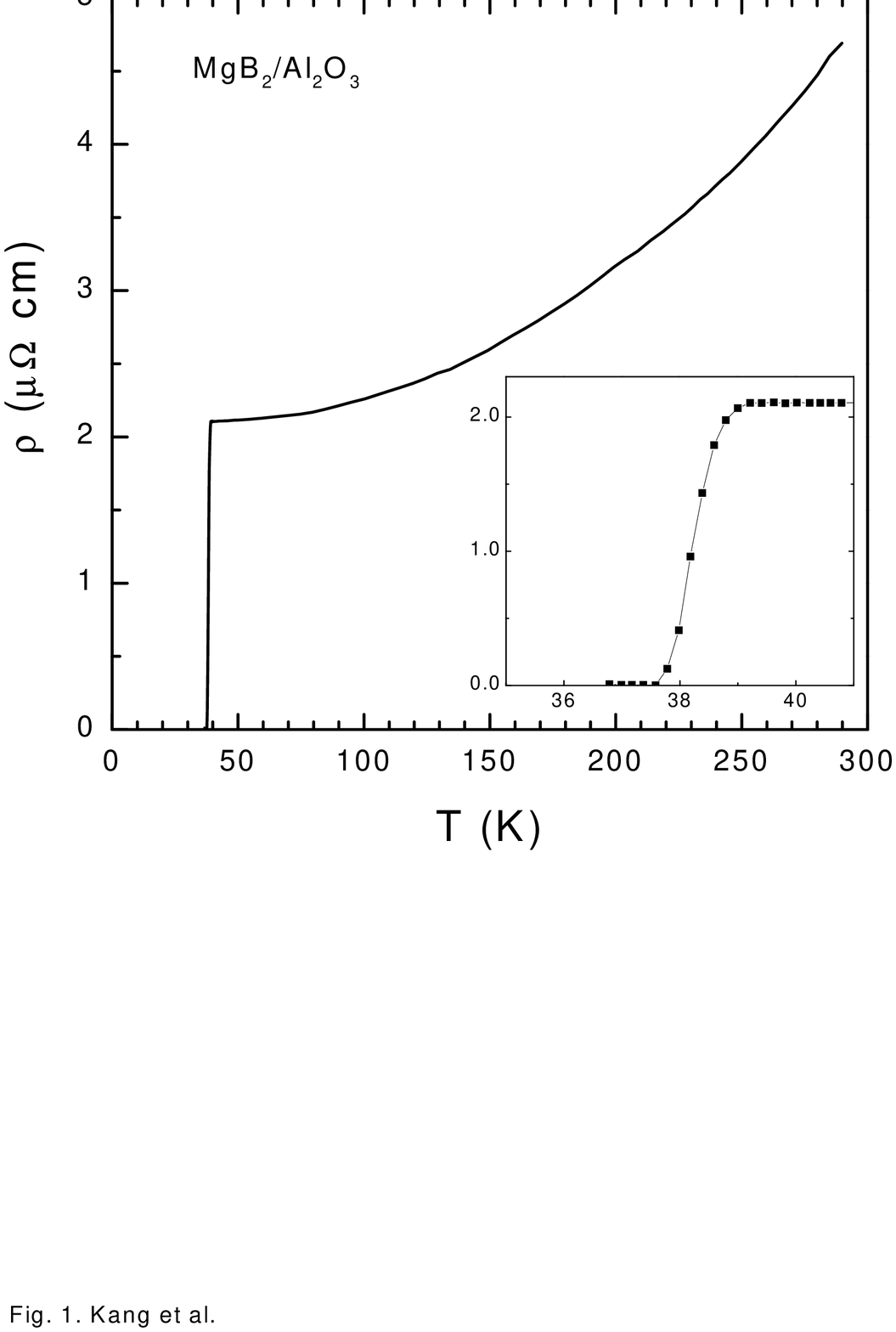, width=6.5cm}
\caption{Resistivity vs. temperature for a MgB$_{2}$ thin film grown on an Al%
$_{2}$O$_{3}$ substrate by using pulsed laser deposition with post annealing
techniques. The inset is a magnified view of the temperature region of 36 -
40 K for the sake of clarity. A sharp superconducting transition is observed
at 39 K.}
\label{fig1}
\end{figure}

\begin{figure}[tbp]
\centering \epsfig{file=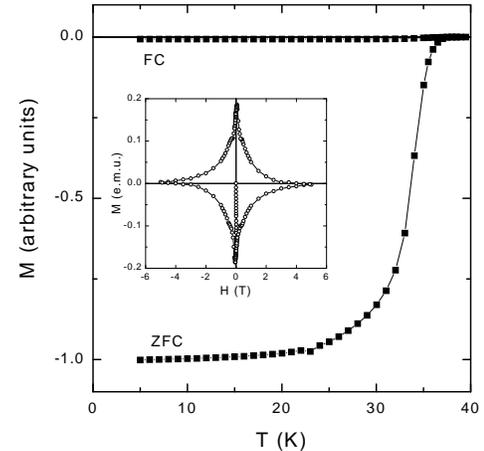, width=6.5cm}
\caption{Magnetization vs. temperature at H = 10 Oe for a
MgB$_{2}$ thin film. The inset shows the M-H hysteresis loop at 5
K. Note that a very high current-carrying capability of $\sim$ 8
x 10$^{6}$ A/cm$^{2}$ was observed at zero field.} \label{fig2}
\end{figure}

\begin{figure}[tbp]
\centering \epsfig{file=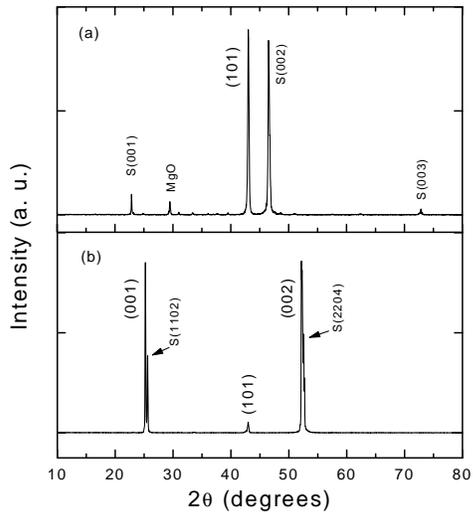, width=6.5cm} \caption{XRD
patterns for MgB$_{2}$ thin films grown on (a) (100) SrTiO$_{3}$
and R-plane (b) (1102) Al$_{2}$O$_{3}$ substrates. The (00{\it
l}) peaks of MgB$_{2}$ grown on Al$_{2}$O$_{3}$ indicate
c-axis-oriented epitaxial thin films whereas MgB$_{2}$ on
SrTiO$_{3}$ show (101) plane-oriented thin films. S denotes the
substrate peaks.}
\label{fig3}
\end{figure}

\end{document}